# From Stoop to Squat:
# A comprehensive analysis of lumbar loading among different lifting styles

Michael VON ARX[1], Melanie LIECHTI[1], Lukas CONNOLLY[2,3,4],
Christian BANGERTER[1], Michael L. MEIER[2,3], Stefan SCHMID[1,5,*]

[1]Bern University of Applied Sciences, Department of Health Professions, Division of Physiotherapy,
Spinal Movement Biomechanics Group, Bern, Switzerland
[2]University of Zurich, Balgrist University Hospital, Department of Chiropractic Medicine,
Integrative Spinal Research, Zurich, Switzerland
[3]University of Zurich, Zurich, Switzerland
[4]ETH Zurich, Department of Health Science and Technology, Zurich, Switzerland
[5]University of Basel, Faculty of Medicine, Basel, Switzerland

**Correspondence:**
PD Dr. Stefan Schmid, Bern University of Applied Sciences, Department of Health Professions,
Murtenstrasse 10, 3008 Bern, Switzerland, +41 79 936 74 79, stefanschmid79@gmail.com

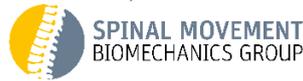

**ABSTRACT**

Lifting up objects from the floor has been identified as a risk factor for low back pain, whereby a flexed spine during lifting is often associated with producing higher loads in the lumbar spine. Even though recent biomechanical studies challenge these assumptions, conclusive evidence is still lacking. This study therefore aimed at comparing lumbar loads among different lifting styles using a comprehensive state-of-the-art motion capture-driven musculoskeletal modeling approach.

Thirty healthy pain-free individuals were enrolled in this study and asked to repetitively lift a 15 kg-box by applying 1) a freestyle, 2) a squat and 3) a stoop lifting technique. Whole-body kinematics were recorded using a 16-camera optical motion capture system and used to drive a full-body musculoskeletal model including a detailed thoracolumbar spine. Continuous as well as peak compressive, anterior-posterior shear and total loads (resultant load vector of the compressive and shear load vectors) were calculated based on a static optimization approach and expressed as factor body weight (BW). In addition, lumbar lordosis angles and total lifting time were calculated. All parameters were compared among the lifting styles using a repeated measures design.

For each lifting style, loads increased towards the caudal end of the lumbar spine. For all lumbar segments, stoop lifting showed significantly lower compressive and total loads (-0.3 to -1.0BW) when compared to freestyle and squat lifting. Stoop lifting produced higher shear loads (+0.1 to +0.8BW) in the segments T12/L1 to L4/L5, but lower loads in L5/S1 (-0.2 to -0.4BW). Peak compressive and total loads during squat lifting occurred approximately 30% earlier in the lifting cycle compared to stoop lifting. Stoop lifting showed larger lumbar lordosis range of motion (35.9±10.1°) than freestyle (24.2±7.3°) and squat (25.1±8.2°) lifting. Lifting time differed significantly with freestyle being executed the fastest (4.6±0.7s), followed by squat (4.9±0.7s) and stoop (5.9±1.1s).

Stoop lifting produced lower total and compressive lumbar loads than squat lifting. Shear loads were generally higher during stoop lifting, except for the L5/S1 segment, where anterior shear loads were higher during squat lifting. Lifting time was identified as another important factor, considering that slower speeds seem to result in lower loads.

# 1. INTRODUCTION

The importance of the correct lifting posture is believed to be strongly connected to the prevention of low back pain (LBP) (Balagué et al., 2012; Caneiro et al., 2019). Even healthcare professionals associate a flexed spine during lifting with danger and therefore seem to influence how people lift every day (Nolan et al., 2018). While lifting has been identified as a main risk factor for LBP, research fails to establish a clear connection between LBP, lifting posture and danger to the spine (Balagué et al., 2012; Saraceni et al., 2020; Schaafsma et al., 2015; Van Dieën et al., 1999). It is widely believed that a flexed spine causes higher spinal loads that could result in structural damage or lead to back complaints in the long-term. Furthermore, the interaction between shear and compressive loads and spine tolerance is still poorly understood (Bazrgari et al., 2007; Gallagher and Marras, 2012), and many of the assumptions regarding load tolerances of the spine are solely based on in vitro studies (Gallagher and Marras, 2012).

Van Dieën et al. (1999) concluded in their review that there was not enough evidence to support advocating the squat technique as a means of preventing LBP. In addition, more recent research suggests that differences in spinal loads among various lifting styles are relatively small and a straight back (spine in a neutral position) might not always be the optimal position (Dreischarf et al., 2016; Khoddam-Khorasani et al., 2020; Kingma et al., 2010; van der Have et al., 2019; Wang et al., 2012). Some suggest that a single optimal position for all situations does not exist (Burgess-Limerick, 2003) and that the lifting technique should be adapted to the lifted weight (Wang et al., 2012). Despite these facts, however, squat lifting still remains the recommended technique (Bazrgari and Shirazi-Adl, 2007; van der Have et al., 2019), which spurs a call for more comprehensive investigations of spinal loading during lifting.

Motion capture-driven musculoskeletal spine modeling is a reliable and non-invasive analysis tool, which allows the calculation of spinal loads in an environment close to the natural movement of the spine. However, many of the available models are highly simplified by using lumped segment models or generic spinal alignments, which limits the accuracy for simulating intersegmental spinal loading during functional activities. To overcome such shortcomings, Connolly et al. (2021) recently introduced a novel approach for modeling subject-specific spinal alignment based on the external back profile obtained from skin marker-based motion capture data, allowing simulations of spinal loading using models with fully articulated thoracolumbar spines.

Furthermore, the currently available studies investigating spinal loading during object lifting solely focused on the analysis of predetermined discrete parameters such as peak forces and none of them included quantitative analyses of data over time. Using such 0-dimensional scalar parameters means that only particular instances of the measurement domain are taken into account, whereby differences during other instances along the time dimension might be missed (regional focus bias) (Papi et al., 2020). To address these issues, Statistical Parametric Mapping (SPM) can be applied (Pataky et al., 2013) which uses Random Field Theory (Adler, 2007) to identify statistical interference over 1-dimensional continuous vectors.

For these reasons, this study aimed at comparing compressive, anterior-posterior shear and total loads of the lumbar spine between freestyle, squat and stoop lifting using a novel subject-specific musculoskeletal modeling approach of the spine as well as advanced statistical methods for analyzing continuous data. Furthermore, lumbar lordosis angles as well as lifting movement duration were investigated for supporting the interpretation of the loads. Such comprehensive knowledge might help to shed more light into the question of how different lifting techniques affect spinal loading.

## 2. MATERIALS AND METHODS

### 2.1. Study population

Thirty healthy pain-free adults (20 males and 10 females; age: 31.8 ± 8.5 years; body height: 175.3 ± 7.5 cm; body mass: 71.7 ± 10.2 kg; BMI: 23.3 ± 2.4 kg/m$^2$; sporting activities per week: 5.3 ± 4.3 hours) were included in this cross-sectional, observational study. Recruitment took place in the personal and workplace environment of the investigators. Inclusion criteria were: aged between 18 and 65 years, ability to perform the required lifting tasks as well as sufficient understanding of the German language. Individuals were excluded in case of any history of LBP in the past 6 months, injuries or operations on the spine, hip, knee or ankle as well as any comorbidities or circumstances (e.g., pregnancy) that could limit the lifting capabilities. In addition, weightlifters, CrossFit athletes, physical therapists and nurses were not eligible due to a potential bias regarding lifting techniques. The study protocol was evaluated by the local ethics committee (Kantonale Ethikkommission Bern, Req-2020-00364) and all participants provided written informed consent prior to collecting any personal or health related data.

### 2.2. Data collection

#### 2.2.1. Subject preparation and instrumentation

Data collection procedures were defined in a detailed case report form (CRF) and carried out in the same manner for each subject by the same two experienced physical therapists. Socio-economic and biometric information such as profession and physical activity level as well as age, sex, body mass and body height were collected prior to any biomechanical measurements. Subsequently, participants were equipped with 58 retro-reflective markers according to the configuration described by Schmid et al. (2017) (Figure 1).

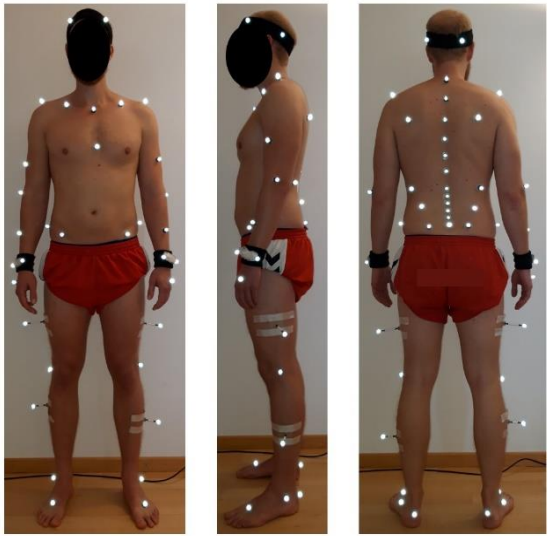

**Figure 1.** Full marker set equipped according to the IfB trunk model in combination with the Plug-in Gait full body model as described in Schmid et al. (2017).

To enable detailed tracking of spinal motion, the configuration included markers placed on the spinous processes of the vertebrae C7, T3, T5, T7, T9, T11, L1-L5 and the sacrum (S1). Two additional markers were placed on the box to allow for adequate detection of the start of the lifting-up movement. Kinematic data were recorded using a 16-camera optical motion capture system (Vicon, Oxford, UK; sampling frequency: 200 Hz). In addition, ground reaction forces were recorded using an embedded force plate (AMTI BP400600, Advanced medical technology Inc., Watertown, MA, USA).

*2.2.2. Lifting tasks*

Subjects were asked to repetitively lift up a 15 kg-box from the floor using a 1) freestyle, 2) squat and 3) stoop lifting technique (Figure 2). Participants were given up to five minutes of practice time until the execution of the lifting technique matched the investigators demands.

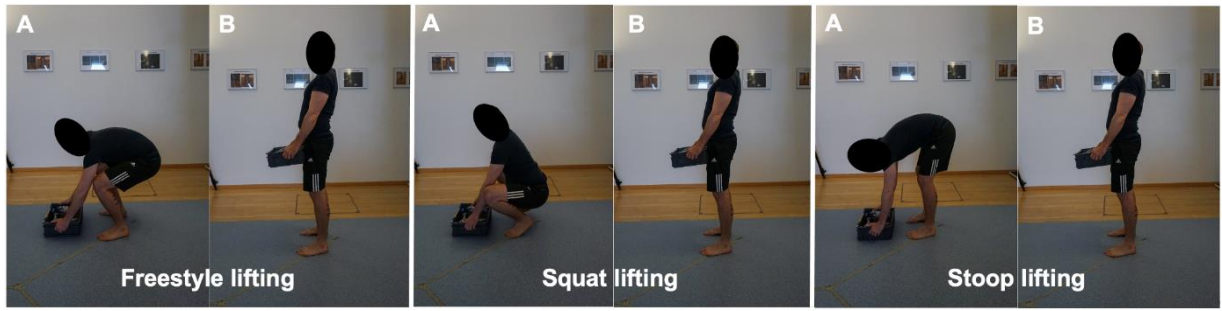

**Figure 2.** Start (A) and end positions (B) of a lifting-up cycle for all three styles. The section of interest spanned from the moment the box left the floor until the subject reached a stable upright standing position.

For each lifting style, subjects had to perform five valid repetitions. A number of key characteristics were defined for each lifting technique, which were closely observed by the investigators during each repetition. All three lifting styles started with the feet parallel about hip width apart and 15 cm behind the box. The box on the floor had to be grabbed with both hands and be placed back on the same place. Start and end of one repetition were standing upright without the box in hands. Only the lifting up sections were used for analysis.

Instructions for freestyle lifting were simply to lift the box in the most comfortable manner, while keeping the feet in place and grabbing the box with both hands. Instructions for squat and stoop lifting were based on Dreischarf et al. (2016). Squat lifting was thereby characterized as lifting with the back kept as straight as possible and with mainly flexing the knees and the hips. Participants were asked to keep the feet flat on the ground if possible. If ankle mobility was insufficient for keeping the feet flat, subjects were tolerated to raise their heels and to stand on the forefoot in order to comply with the instruction of keeping the back as straight as possible. Stoop lifting was characterized by bending forward with a clear flexion of the spine and with the knees kept as straight as possible while bending in the hips. Subjects that were able to perform this lift with a straight back and straight legs by solely flexing in the hips were reminded to clearly flex their lumbar spine for the lift to count as valid.

## 2.3. Data reduction

Data was pre-processed with the Nexus software (version 2.6, Vicon UK, Oxford, UK), which included the reconstruction and labeling of the markers as well as filtering of the trajectories. Additionally, temporal events were manually set to identify the sections of interest, i.e. the sections containing the lifting up movements. For detection of the exact start and end points, a custom MATLAB routine (R2020b; MathWorks, Inc., Natick Massachusetts, USA) was used. In brief, the start of the movement was defined as the point where the vertical velocity of the C7 marker initially exceeded 5% of the maximal vertical velocity, and the end of the movement was defined as the point where the vertical velocity fell below this 5% threshold (Connolly et al., 2021).

For determining spinal loading, we used previously introduced OpenSim-based female and male musculoskeletal full-body models including a detailed and fully articulated thoracolumbar spine (Connolly et al., 2021) (Figure 3). To enable subject-specific simulations, we used the OpenSim Scaling Tool to scale segment lengths and masses based on the marker data and total body mass, respectively. In addition, a custom MATLAB algorithm was applied to adjust the sagittal plane spinal curvatures based on the markers placed on the spinous processes, the head and the sacrum (Connolly et al., 2021). Simulations were driven by kinematic (derived from the marker data using the OpenSim Inverse Kinematics Tool) and ground reaction force data and solved using static optimization with a cost function that minimizes the sum of squared muscle activation (Herzog, 1987). Intersegmental joint forces were computed using OpenSim Joint Reaction Analysis.

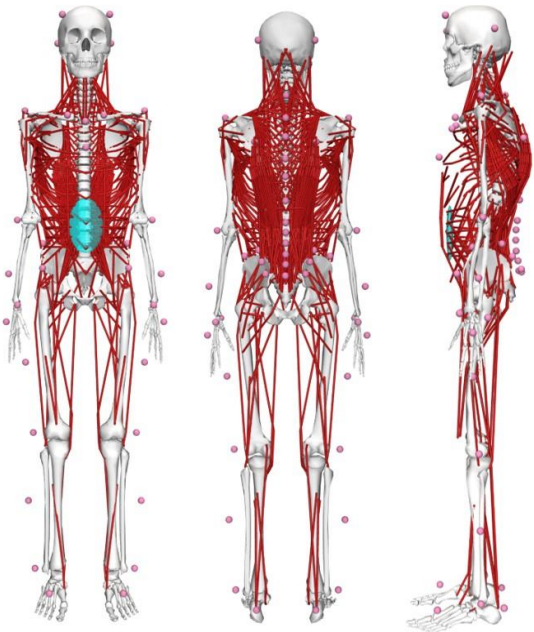

**Figure 3.** OpenSim-based musculoskeletal full-body models including a detailed and fully articulated thoracolumbar spine and 58 virtual skin markers to allow for subject-specific model scaling as well as comprehensive simulation of spinal loading during dynamic functional activities using motion capture data.

Lumbar lordosis angles were calculated using a custom MATLAB routine as described in Schmid et al. (2017). In brief, we applied a combination of a quadratic polynomial and a circle fit function to the sagittal plane trajectories of the markers placed on the spinous processes of L1-S1 and used the central angle theorem to calculate the lumbar lordosis angle. Primary outcome variables were continuous as well as peak compressive forces, anterior-posterior (AP) shear forces and total forces (resultant force vector of the compressive and AP shear force vectors) for the lumbar segments L1/2 to L5/S1 (expressed as a factor of body weight [BW]). Secondary outcome variables included lumbar lordosis angle range of motion (RoM; expressed in degrees) as well as lifting movement duration (time between start and end points of lifting-up phase, expressed as dimensionless number according to Hof (1996)).

## 2.4. Statistical analysis

Statistical analysis was performed using MATLAB with the package 'spm1d' for one-dimensional Statistical Parametric Mapping (SPM; www.spm1d.org) for continuous data and RStudio (version 1.3.1093, R foundation for statistical computing, Vienna, Austria) for discrete parameters. Normal distribution was verified using the SPM-function 'spm1d.stats.normality.anova1rm' for continuous data and the Shapiro Wilk test and Q-Q-plot inspection for discrete parameters. Differences among the three lifting styles were investigated using the SPM-functions 'spm1d.stats.anova1rm' and 'spm1d.stats.ttest_paired' for continuous data as well as repeated measures analyses of variance (ANOVA) with paired t-tests for post hoc analyses for discrete parameters. The alpha level was set at 0.05 for the ANOVAs and 0.017 (Bonferroni-corrected) for the post hoc tests.

## 3. RESULTS

For three participants, musculoskeletal simulations were not conducted due to insufficient marker recognition in the anterior thorax region, leaving a sample of 27 for the spinal loading parameters. The calculation of lumbar lordosis angle and lifting movement duration, on the other hand, was based on all 30 participants. Means and standard deviations as well as p-values of the statistical analyses for the continuous and peak spinal loads can be found in the Supplementary Material.

### 3.1. Continuous loads

ANOVAs showed significant differences between lifting styles for all segments and load types. Results showed increasing loads towards the caudal end of the lumbar spine for all styles (Figures 4-6). Significant differences between styles occurred predominantly during the first 50% of the lifting cycle and got smaller towards the end of the cycle.

The analysis of total and compressive loads revealed that stoop lifting produced significantly smaller loads compared to both other styles in all segments and that the loads for freestyle and squat lifting were mostly similar, with only few differences in the L4/L5 and L5/S1 segments for short sections of the lifting movement (Figures 4 and 5). Moreover, the onset of peak total loading occurred later in the cycle for stoop lifting when compared to squat and freestyle lifting.

AP shear loads analysis showed significant differences between all styles in all lumbar segments (Figure 6). Stoop lifting produced generally higher shear loads, except in the L5/S1 segment, where shear forces were smaller compared to the other lifting styles.

### 3.2. Peak loads

ANOVAs showed significant differences between lifting styles for all segments and load types. For all styles and all three load types, peak loads increased towards the caudal end of the spine with the largest loads occurring in the L5/S1 segment (Figures 7 to 9). In addition, there was a trend for smaller differences in compressive and peak loads between styles towards the lower end of the spine, indicating that differences between styles are more pronounced in the upper part of the lumbar spine.

Peak total and compressive loads for stoop lifting were significantly smaller in every segment, when compared to both other styles (Figures 7 and 8). No significant differences in total and compressive loads were found between squat and freestyle lifting in the segments T12/L1 to L2/L3, while in the segments L3/L4 to L5/S1, freestyle produced significantly larger loads than both other styles.

Peak AP shear loads in the L5/S1 segment for all styles were up to 23 times larger as in the other segments (Figure 9). Stoop lifting resulted in significantly larger shear loads throughout the lumbar spine, except for the segment L5/S1. For the segments T12/L1 to L4/L5, squat lifting produced significantly smaller shear loads than both other styles.

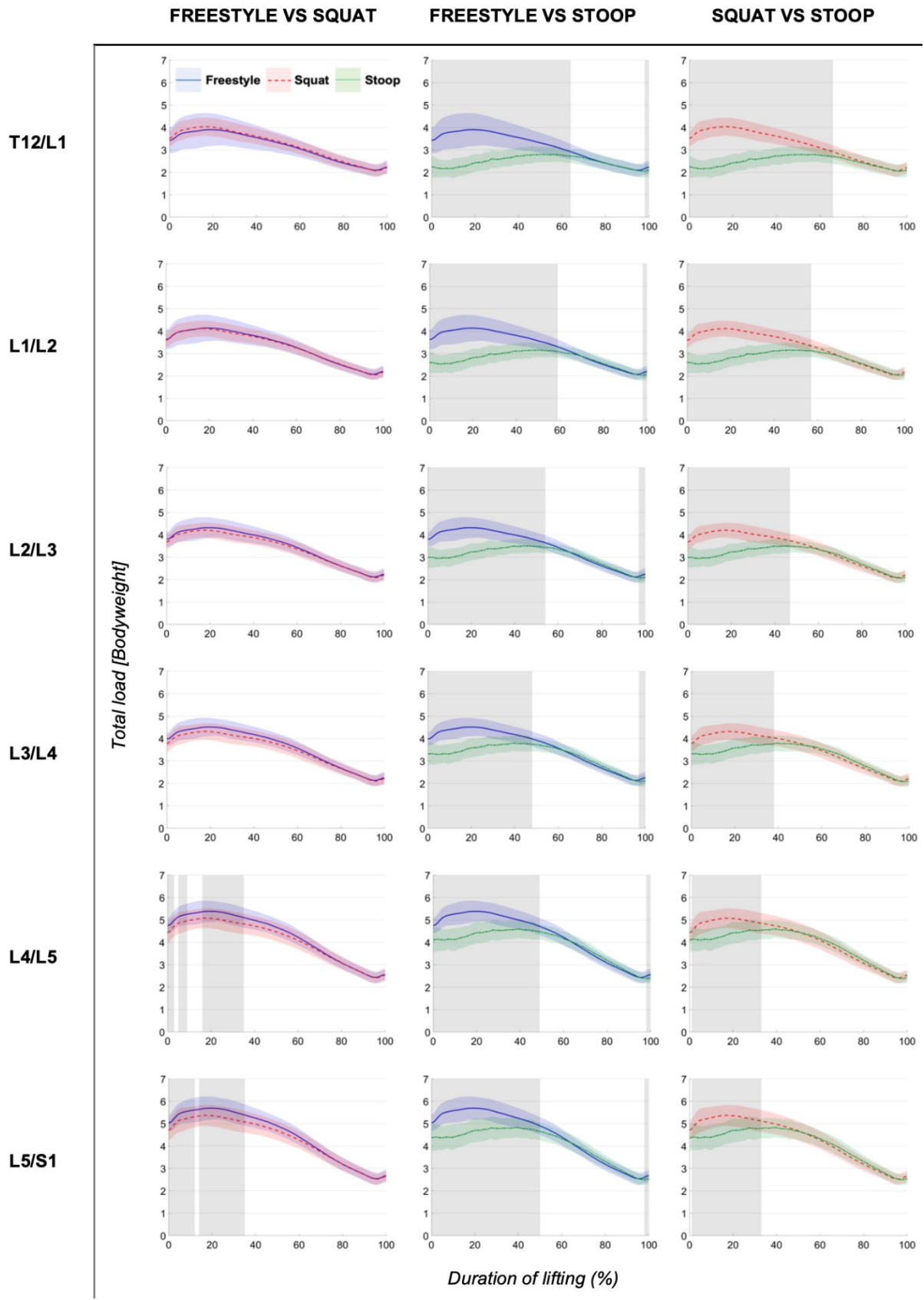

**Figure 4.** Continuous total loads graphs arranged by compared styles and spinal segment. Curves depict the respective total loads throughout the lift up cycle, starting when the box leaves the ground (0%) to upright standing position (100%). Colored areas above and below the curves indicate the standard deviation and the greyed sectors in the graphs indicate the parts of the lifting cycle where significant differences between styles were detected.

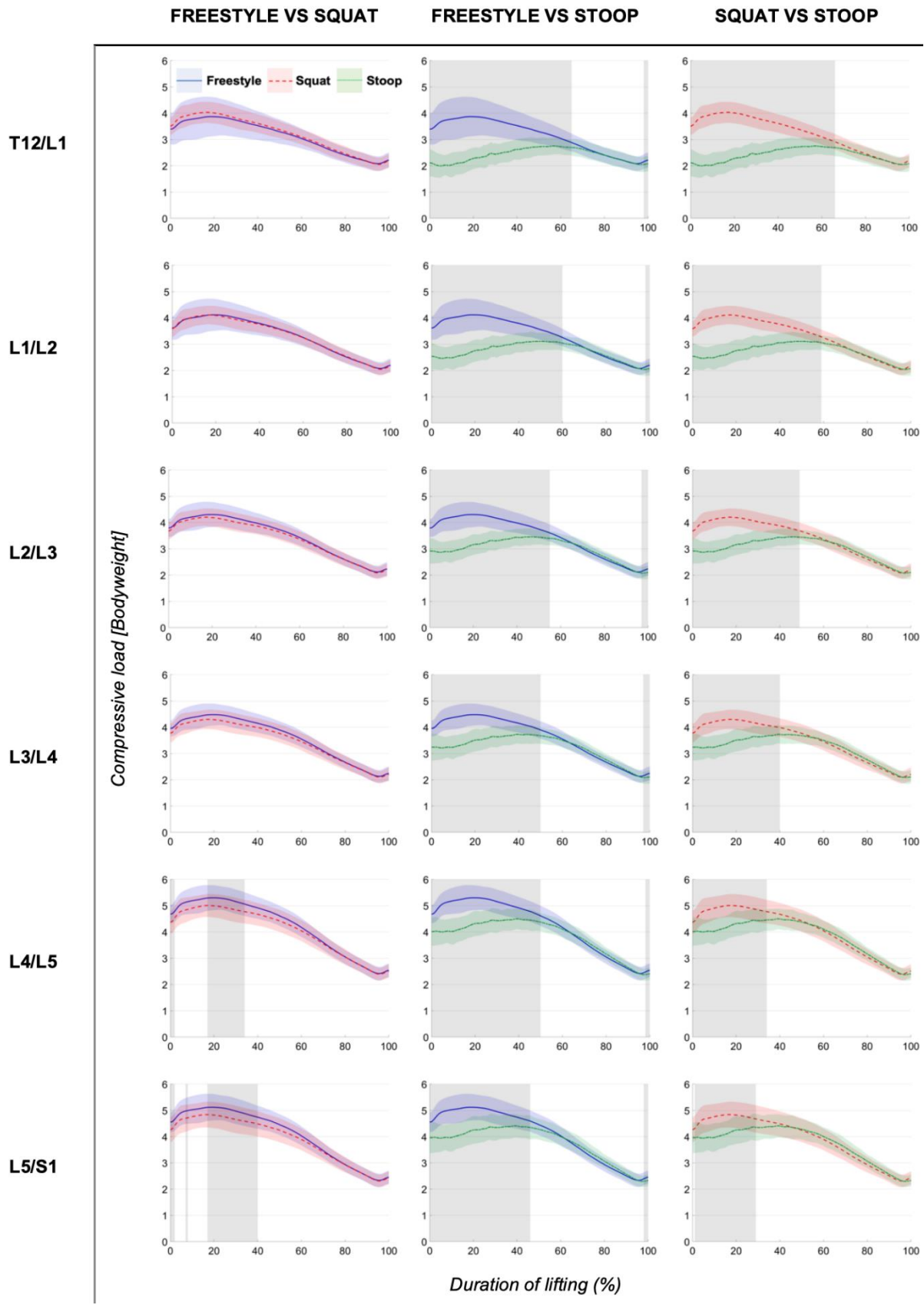

**Figure 5.** Continuous compressive loads graphs arranged by compared styles and spinal segment. Curves depict the respective compressive loads throughout the lift up cycle, starting when the box leaves the ground (0%) to upright standing position (100%). Colored areas above and below the curves indicate the standard deviation and the greyed sectors in the graphs indicate the parts of the lifting cycle where significant differences between styles were detected.

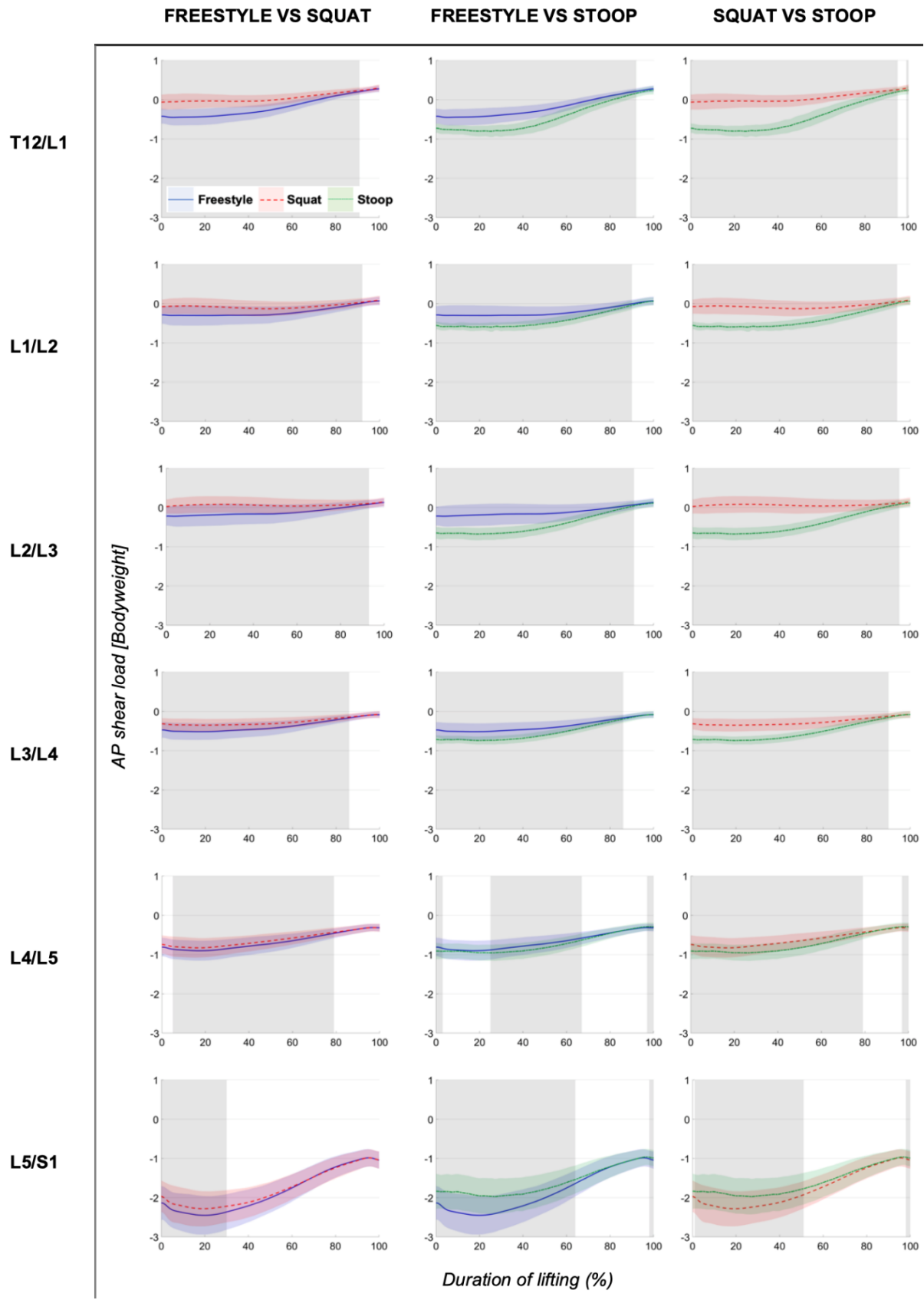

**Figure 6.** Continuous AP shear loads graphs arranged by compared styles and spinal segment. Curves depict the respective AP shear loads throughout the lift up cycle, starting when the box leaves the ground (0%) to upright standing position (100%). Colored areas above and below the curves indicate the standard deviation and the greyed sectors in the graphs indicate the parts of the lifting cycle where significant differences between styles were detected.

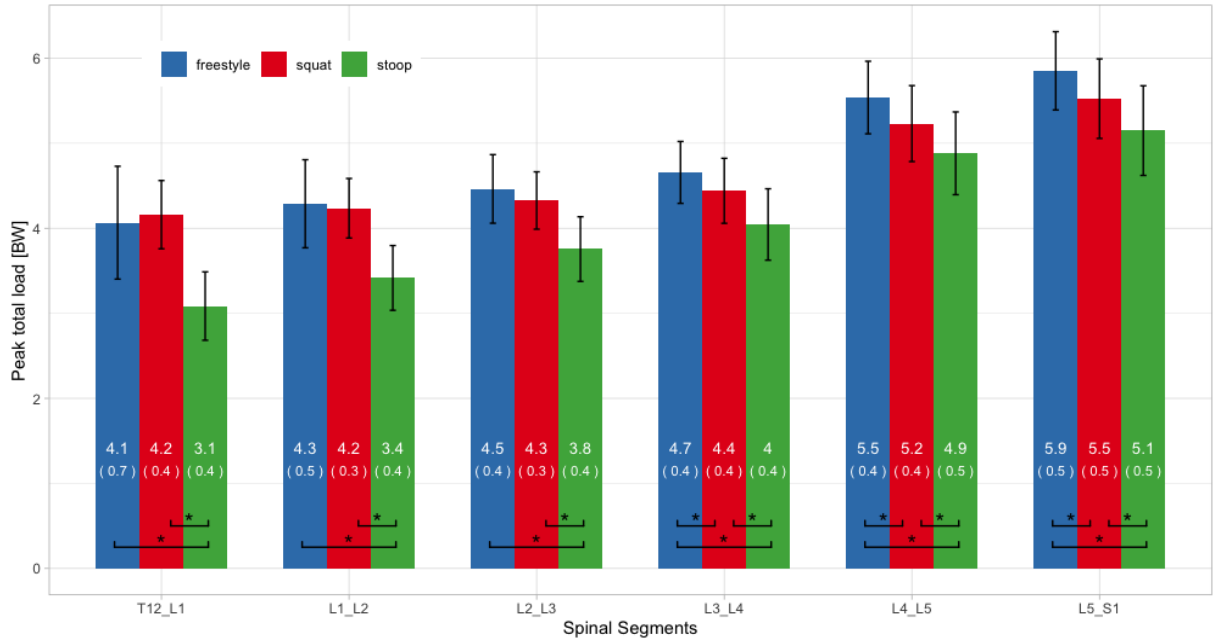

**Figure 7.** Peak total loads of all three lifting styles grouped by spinal segments. Bars represent the mean loads normalized to bodyweight (BW). Mean and standard deviation values are listed in the bar centers. Horizontal parentheses at the bottom of bar groups indicate comparisons for which a significant difference (*) was detected in the post hoc analysis. Lines at the bar ends indicate standard deviation.

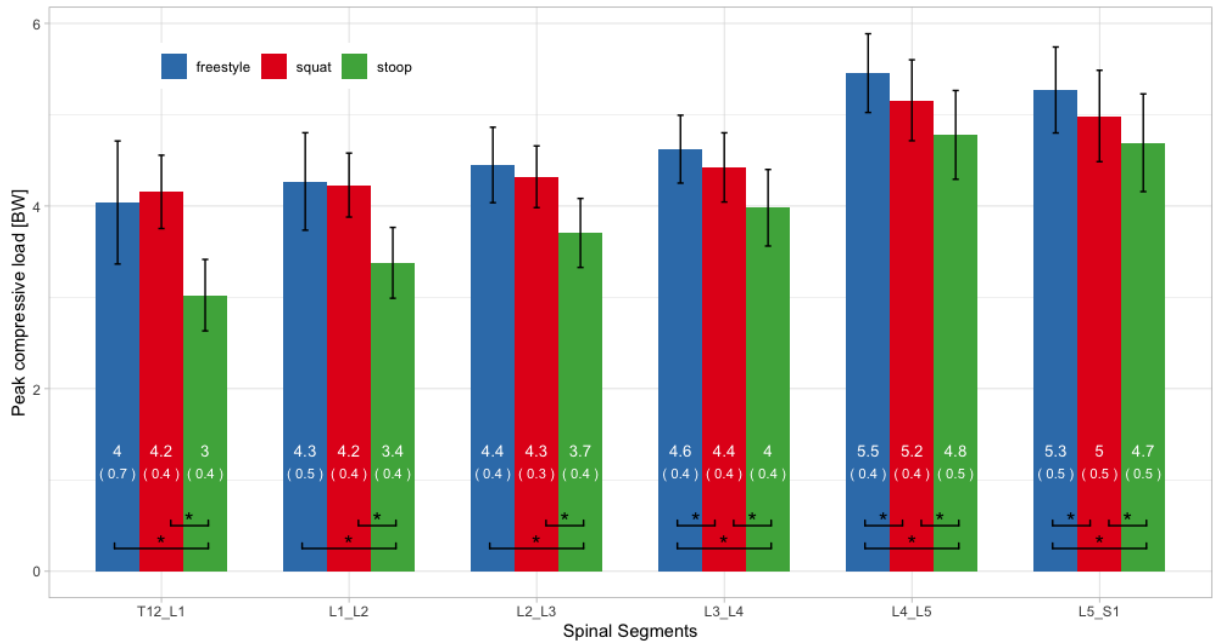

**Figure 8.** Peak compressive loads of all three lifting styles grouped by spinal segments. Bars represent the mean loads normalized to bodyweight (BW). Mean and standard deviation values are listed in the bar centers. Horizontal parentheses at the bottom of bar groups indicate comparisons for which a significant difference (*) was detected in the post hoc analysis. Lines at the bar ends indicate standard deviation.

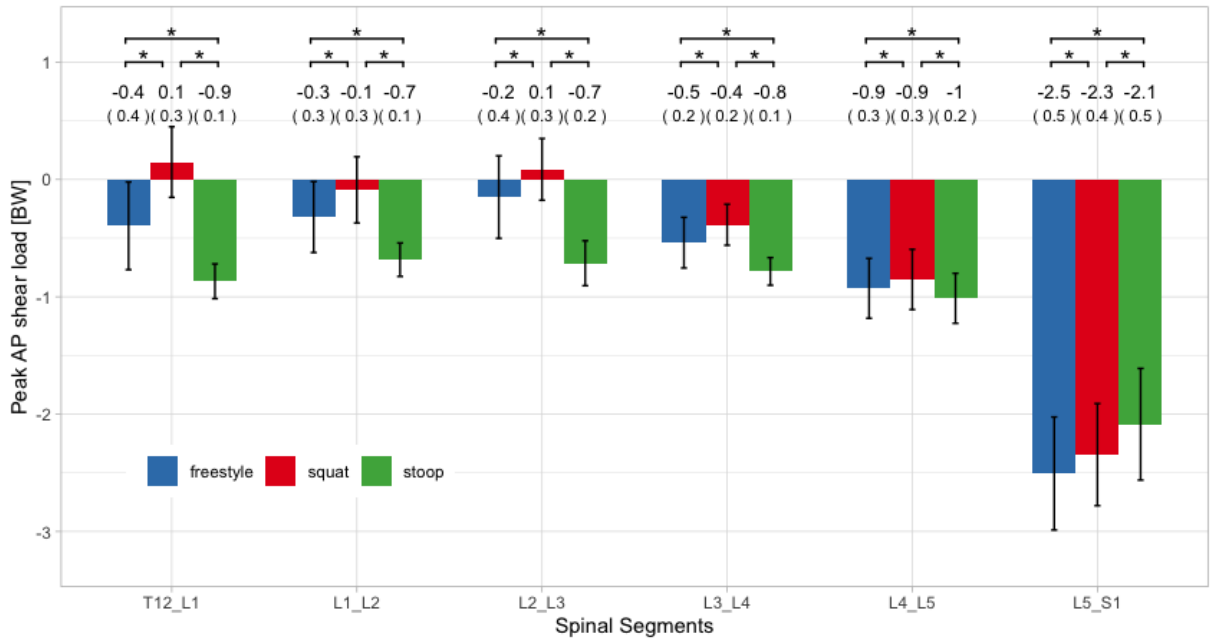

**Figure 9.** Peak AP shear loads of all three lifting styles grouped by spinal segments. Bars represent the mean loads normalized to bodyweight (BW). Mean and standard deviation values are listed above the bars. Horizontal parentheses above bar groups indicate comparisons for which a significant difference (*) was detected in the post hoc analysis. Lines at the bar ends indicate standard deviation.

### 3.3. Lumbar lordosis angle RoM and lifting movement duration

The analysis of the lumbar lordosis angle RoM showed mean values of 24.2 ± 7.3 degrees for freestyle, 25.1 ± 8.2 degrees for squat and 35.9 ± 10.1 degrees for stoop lifting. ANOVA revealed significant differences between styles (p<0.001). Post hoc analysis revealed significant differences between stoop and squat lifting (p<0.001) as well as between stoop and freestyle lifting (p<0.001). No significant difference was found between squat and freestyle lifting.

Regarding lifting movement duration, freestyle lifting was performed the fastest with a mean duration of 4.6 ± 0.7, followed by squat lifting with 4.9 ± 0.7 and stoop lifting with 5.9 ± 1.1. The statistical analysis indicated significant differences between freestyle and squat lifting (p=0.004), freestyle and stoop lifting (p<0.001) as well as squat and stoop lifting (p<0.001). Additional analyses showed trends for negative relationships between spinal loads and lifting movement duration (see Supplementary Material).

### 4. DISCUSSION

This study aimed at exploring differences in lumbar spine loading between freestyle, squat and stoop lifting using a comprehensive motion capture-driven musculoskeletal full-body modeling approach. Results demonstrated that stoop lifting produced smaller total and compressive loads compared to squat lifting. Moreover, stoop lifting generally resulted in higher AP shear loads, except for the L5/S1 segment, where AP shear loads were the smallest compared to the other lifting styles.

The fact that stoop lifting produced smaller compressive loads is consistent with Khoddam-Khorasani et al. (2020); Kingma et al. (2004); Potvin et al. (1991) and Leskinen et al. (1983). On the other hand, the findings partially disagree with Bazrgari et al. (2007), Anderson and Chaffin (1986) and Faber et al. (2009), who found that stoop lifting resulted in larger L5/S1 compressive loads than squat lifting. Furthermore, Dreischarf et al. (2016); Hwang et al.

(2009); Kingma et al. (2010) and Troup et al. (1983) reported no significant difference in spinal compression between squat and stoop lifting. Reasons for such inconsistent findings could be differences in the experimental settings as well as the underlying models. Changes in lifting style execution, variations in lowering depth or horizontal distance of the weight to S1 might considerably influence spinal loading. This issue was also mentioned by Van Dieën et al. (1999) and could be addressed in the future with better standardization in the experimental designs.

While compressive loads in this study were up to 43 times larger than shear loads, shear forces are still a subject of great interest. Gallagher and Marras (2012) reported that especially spines of younger individuals (less than 30 years) might be particularly susceptible to shear loads due to higher disc elasticity. For all lifting styles evaluated in this study, AP shear loads reached magnitudes of about 2.5 BW in the L5/S1 segment, which was consistent with Khoddam-Khorasani et al. (2020); Kingma et al. (2004) and Bazrgari et al. (2007). The 180% increase in peak L4/L5 shear load during stoop compared to squat lifting reported by Potvin et al. (1991) was not reproduced in our experiment. Nonetheless, our simulations showed shear load increases ranging from 100% (L3/L4) to 800% (T12/L1) in segments above L4/L5. No significant differences in L5/S1 shear loads between stoop and squat lifting were reported by Kingma et al. (2004) and Kingma et al. (2010). In this study, L5/S1 was the only segment where shear loads were larger during squat compared to stoop lifting (about 10%). In contrast, Bazrgari et al. (2007) found larger shear for this segment during stoop lifting. Shear forces appear to be highly dependent on the model used (Van Dieën et al., 1999). Kingma et al. (2004) explained the lack of significant differences between lifting styles with a high between-subject variance of the shear forces. Reasons for such differing results could be different horizontal distances of the lifted weight to S1, different lumbar flexion angles or other confounding variables such as variations in lifting style execution or differences in starting positions (grip height).

Potvin et al. (1991) suggested, that shear loads are more strongly influenced by lumbar flexion angles than lifted weight. Compressive loads behave differently in this aspect as they increase linearly with added weight (Marras et al., 1999; Potvin et al., 1991). This would imply that lumbar flexion angles are a confounding variable when comparing shear loads, if not controlled for.

In this study, freestyle lifting generated larger spinal loads than squat lifting. This agrees with results of Kingma et al. (2010) where freestyle produced larger peak L5/S1 compression and shear forces than squat or stoop, although differences were not statistically significant. Moreover, Dolan et al. (1994b) reported that freestyle lifting generated larger net moments than both other styles but suspected this result to be mainly due to a faster execution of the freestyle lifts. In the studies by (Kingma et al., 2004) and (Khoddam-Khorasani et al., 2020), spinal loads during freestyle lifting fell in between those during squat and stoop lifting. Reason for these differences could be the variations in the experimental setting or the used models. In our study and the study conducted by Kingma et al. (2010) participants lifted a box from the floor, while in the study by Khoddam-Khorasani et al. (2020) participants were measured in isometrically held positions of 40° and 65° forward upper trunk inclination with and without holding a weight.

While loads increased for all lifting styles towards the caudal end of the lumbar spine, differences between lifting styles seemed more pronounced in the upper lumbar spine. Similar results were found by Khoddam-Khorasani et al. (2020), suggesting that differences between lifting styles become less relevant towards the caudal end of the spine.

Time related analysis revealed that peak loads occur at different time segments for squat lifting and stoop lifting. During squat lifting, the highest loads occurred within the first 30% of the lifting cycle, whereas during stoop lifting, peak loads were indicated between 40% and 70% of the lifting cycle. Faber et al. (2009) reported an early onset of peak loading but did not differentiate further between styles or within the lifting cycle. Referring to the strain rate dependency of vertebral discs (Kemper et al., 2007), a slower onset of peak loading during stoop lifting might result in less stress on the spine.

It has to be considered that at least a part of the differences in spinal loading between the lifting styles might have been due to differences in lifting movement duration. Stoop lifting was executed about 20% slower than squat lifting and about 30% slower than freestyle lifting. These slower lifting speeds are consistent with the findings of van der Have et al. (2019) but not with those of Straker (2003), who stated that stoop lifting is generally performed faster and is therefore less fatiguing than squat lifting. Trunk movement speed was shown to have a direct influence on spinal loading (Bazrgari et al., 2007; Dolan et al., 1994a). Faster lifting speeds thereby lead to larger net moments, suggesting that dynamic factors might have a larger impact on spinal loading than lifting technique (Kjellberg et al., 1998). Frost et al. (2015) demonstrated that movement strategies change when the same task is repeated with different speeds. van der Have et al. (2019) therefore suggested that faster lifting speeds should be favored as it might reduce muscle fatigue.

The lumbar lordosis angle RoMs measured in this study are consistent with previously reported findings (Kingma et al., 2004; Kingma et al., 2010; Potvin et al., 1991). Although RoM angles were smaller during squat lifting compared to stoop lifting, there is a considerable amount of lumbar flexion occurring even when specifically asked to keep a straight back. Pavlova et al. (2018) even suggested that individuals alter their lifting style primarily by altering knee joint flexion, while retaining similar lumbar spine motion as during freestyle lifting. Nevertheless, the fact that the spine never stays truly neutral when lifting should be kept in mind when discussing lumbar posture and lifting.

Limitations of this study include the specific biometric profile of the test group (age, fitness level and sex distribution), which makes the results not transferrable to a general population. Further, not randomizing the sequence of lifting styles can be considered another limitation. Since stoop lifting was always performed last, fatigue might have resulted in stoop lifting being executed the slowest.

Future research should include broadening the demographic and biometric parameters and include more diverse sample groups or explore lumbar loads among different lifting styles in combination with different lifting speeds. In addition, weights might be adjusted to individual strength levels of the participants. Kingma et al. (2010) reported that when using a 15 kg weight, the impact of trunk inclination outweighed the influence of the weight. In this experiment some subjects reported that the 15 kg box felt heavy, while others considered it light. Increasing the weight close to a subject's individual maximum should pronounce the effect of weight in relation to trunk inclination. Another topic for further research could be the interaction of shear loads in relation to different lumbar flexion angles and different weights.

The reason why squat lifting often remains the recommended lifting technique seems to come down to other factors than just spinal loading such as muscle fatigue or the sensitivity of passive properties of the spine (Bazrgari and Shirazi-Adl, 2007; van der Have et al., 2019). Based on the fatigue-failure-theorem (Gallagher and Heberger, 2013; Gallagher and Schall, 2017) future research should consider the duration of lifting in the risk assessment (van der Have et al., 2019). However, for single repetitions and moderate weights, recommendations should be reevaluated.

In conclusion, this work showed that stoop lifting produced lower total and compressive lumbar loads than squat lifting. Shear loads were generally higher during stoop lifting, except for the L5/S1 segment, where anterior shear loads were higher during squat lifting. While loads consistently increased towards the lower end of the spine, differences in spinal loading between lifting styles were more pronounced in the upper part of the lumbar spine. Considering that freestyle lifting was executed the fastest and stoop lifting the slowest, the differences in spinal loads might have partially been influenced by different lifting speeds.

Additionally, the clearly noticeable lumbar spinal flexion occurring during squat lifting suggests that the spine never stays fully neutral during lifting, even when specifically asked to not flex the spine. The findings of this study provide further support to the notion that there is no one-size-fits-all approach and that current lifting guidelines should be reevaluated.

## 5. CONFLICT OF INTEREST STATEMENT

The authors declare that the research was conducted in the absence of any commercial or financial relationships that could be construed as a potential conflict of interest.

## 6. FUNDING

None.

## 7. ACKNOWLEDGMENTS

The authors thank all the volunteers for participating in this study.

# 9. SUPPLEMENTARY MATERIAL

**Supplementary Table 1:** Results from the statistical analysis of continuous loads. Presented are F-Threshold (F), number of clusters (n) and cluster location (location) throughout the lift cycle and p-values for each cluster from the post hoc analysis with SPM. Non-significant results are not shown and left empty (-)

| LOAD | FREESTYLE VS SQUAT | | | | FREESTYLE VS STOOP | | | | SQUAT VS STOOP | | | |
|---|---|---|---|---|---|---|---|---|---|---|---|---|
| TOTAL | F | n | Location | p-Values | F | n | Location | p-Values | F | n | Location | p-Values |
| T12_L1 | - | - | - | - | 3.76 | 2 | 0-64% 98-100% | p=0 p=0.015 | 3.82 | 1 | 0-66% | p=0 |
| L1_L2 | - | - | - | - | 3.78 | 2 | 0-59% 98-100% | p=0 p=0.014 | 3.81 | 1 | 0-57% | p=0 |
| L2_L3 | - | - | - | - | 3.80 | 2 | 0-54% 97-100% | p=0 p=0.013 | 3.81 | 1 | 0-47% | p=0 |
| L3_L4 | - | - | - | - | 3.82 | 2 | 0-48% 97-100% | p=0 p=0.013 | 3.79 | 1 | 0-38% | p=0 |
| L4_L5 | 3.61 | 3 | 0-3% 5-9% 16-35% | p=0.016 p=0.014 p<0.001 | 3.83 | 2 | 0-49% 98-100% | p=0 p=0.014 | 3.79 | 1 | 1-33% | p<0.001 |
| L5_S1 | 3.62 | 2 | 0-12% 14-35% | p=0.003 p<0.001 | 3.84 | 2 | 0-50% 98-100% | p=0 p=0.014 | 3.79 | 1 | 1-33% | p<0.001 |
| COMPRESSIVE | | | | | | | | | | | | |
| T12_L1 | - | - | - | - | 3.77 | 2 | 0-65% 98-100% | p=0 p=0.015 | 3.83 | 1 | 0-66% | p=0 |
| L1_L2 | - | - | - | - | 3.79 | 2 | 0-60% 98-100% | p=0 p=0.014 | 3.82 | 1 | 0-59% | p=0 |
| L2_L3 | - | - | - | - | 3.80 | 2 | 0-55% 97-100% | p=0 p=0.014 | 3.81 | 1 | 0-49% | p=0 |
| L3_L4 | - | - | - | - | 3.81 | 2 | 0-50% 97-100% | p=0 p=0.013 | 3.79 | 1 | 0-40% | p=0 |
| L4_L5 | 3.61 | 3 | 0-2% 6.9-7% 17-34% | p=0.016 p=0.017 p<0.001 | 3.83 | 2 | 0-50% 98-100% | p=0 p=0.014 | 3.79 | 1 | 0-34% | p=0 |
| L5_S1 | 3.62 | 3 | 0-2% 7-8% 17-40% | p=0.016 p=0.017 p<0.001 | 3.85 | 2 | 0-46% 98-100% | p=0 p=0.014 | 3.80 | 1 | 1-29% | p<0.001 |

**Supplementary Table 1:** Cont.

| LOAD | FREESTYLE VS SQUAT | | | | FREESTYLE VS STOOP | | | | SQUAT VS STOOP | | | |
|---|---|---|---|---|---|---|---|---|---|---|---|---|
| **AP SHEAR** | | | | | | | | | | | | |
| T12_L1 | 3.50 | 1 | 0-91% | p=0 | 3.63 | 1 | 0-92% | p=0 | 3.65 | 2 | 0-95% 99-100% | p=0 p=0.017 |
| L1_L2 | 3.48 | 1 | 0-92% | p=0 | 3.63 | 1 | 0-90% | p=0 | 3.65 | 1 | 0-94% | p=0 |
| L2_L3 | 3.44 | 1 | 0-93% | p=0 | 3.46 | 1 | 0-91% | p=0 | 3.47 | 1 | 0-95% | p=0 |
| L3_L4 | 3.46 | 1 | 0-86% | p=0 | 3.52 | 1 | 0-86% | p=0 | 3.54 | 1 | 0-90% | p=0 |
| L4_L5 | 3.56 | 1 | 5-79% | p=0 | 3.69 | 3 | 0-3% 25-67% 97-100% | p=0.014 p<0.001 p=0.015 | 3.64 | 2 | 0-79% 97-100% | p=0 p=0.015 |
| L5_S1 | 3.62 | 1 | 0-30% | p<0.001 | 3.75 | 2 | 0-64% 98-100% | p=0 p=0.015 | 3.68 | 2 | 1-50% 98-100% | p=0 p=0.016 |

**Supplementary Table 2:** Results from the statistical analysis of peak loads. Displays descriptive values (mean, sd) of discrete outcome variables for peak loads (total, compressive and AP shear) per segment and corresponding p-values from ANOVA and post hoc tests. Significance is indicated by *.

| LOAD | FREESTYLE mean (sd) | SQUAT mean (sd) | STOOP mean (sd) | ANOVA | FREESTYLE VS SQUAT | FREESTYLE VS STOOP | SQUAT VS STOOP |
|---|---|---|---|---|---|---|---|
| **TOTAL** | | | | | | | |
| T12_L1 | 4.23 (0.89) | 4.25 (0.50) | 3.08 (0.40) | p<0.001* | p=0.325 | p<0.001* | p<0.001* |
| L1_L2 | 4.43 (0.77) | 4.31 (0.44) | 3.42 (0.38) | p<0.001* | p=0.513 | p<0.001* | p<0.001* |
| L2_L3 | 4.58 (0.64) | 4.38 (0.41) | 3.76 (0.38) | p<0.001* | p=0.062 | p<0.001* | p<0.001* |
| L3_L4 | 4.77 (0.60) | 4.49 (0.43) | 4.04 (0.42) | p<0.001* | p=0.002* | p<0.001* | p<0.001* |
| L4_L5 | 5.64 (0.63) | 5.26 (0.48) | 4.88 (0.49) | p<0.001* | p<0.001* | p<0.001* | p<0.001* |
| L5_S1 | 5.95 (0.61) | 5.55 (0.48) | 5.15 (0.53) | p<0.001* | p<0.001* | p<0.001* | p<0.001* |
| **COMPRESSIVE** | | | | | | | |
| T12_L1 | 4.20 (0.90) | 4.24 (0.50) | 3.02 (0.39) | p<0.001* | p=0.238 | p<0.001* | p<0.001* |
| L1_L2 | 4.42 (0.78) | 4.30 (0.44) | 3.38 (0.39) | p<0.001* | p=0.636 | p<0.001* | p<0.001* |
| L2_L3 | 4.57 (0.65) | 4.37 (0.41) | 3.70 (0.38) | p<0.001* | p=0.084 | p<0.001* | p<0.001* |
| L3_L4 | 4.73 (0.61) | 4.47 (0.43) | 3.98 (0.42) | p<0.001* | p=0.004* | p<0.001* | p<0.001* |
| L4_L5 | 5.56 (0.63) | 5.19 (0.48) | 4.78 (0.49) | p<0.001* | p<0.001* | p<0.001* | p<0.001* |
| L5_S1 | 5.35 (0.61) | 5.00 (0.51) | 4.69 (0.54) | p<0.001* | p<0.001* | p<0.001* | p<0.001* |
| **AP SHEAR** | | | | | | | |
| T12_L1 | -0.40 (0.37) | 0.15 (0.30) | -0.87 (0.15) | p<0.001* | p<0.001* | p<0.001* | p<0.001* |
| L1_L2 | -0.32 (0.30) | -0.09 (0.28) | -0.68 (0.14) | p<0.001* | p<0.001* | p<0.001* | p<0.001* |
| L2_L3 | -0.15 (0.35) | 0.09 (0.26) | -0.71 (0.19) | p<0.001* | p<0.001* | p<0.001* | p<0.001* |
| L3_L4 | -0.54 (0.22) | -0.39 (0.18) | -0.78 (0.12) | p<0.001* | p<0.001* | p<0.001* | p<0.001* |
| L4_L5 | -0.93 (0.26) | -0.85 (0.26) | -1.01 (0.21) | p<0.001* | p<0.001* | p<0.001* | p<0.001* |
| L5_S1 | -2.51 (0.48) | -2.35 (0.44) | -2.09 (0.48) | p<0.001* | p<0.001* | p<0.001* | p<0.001* |

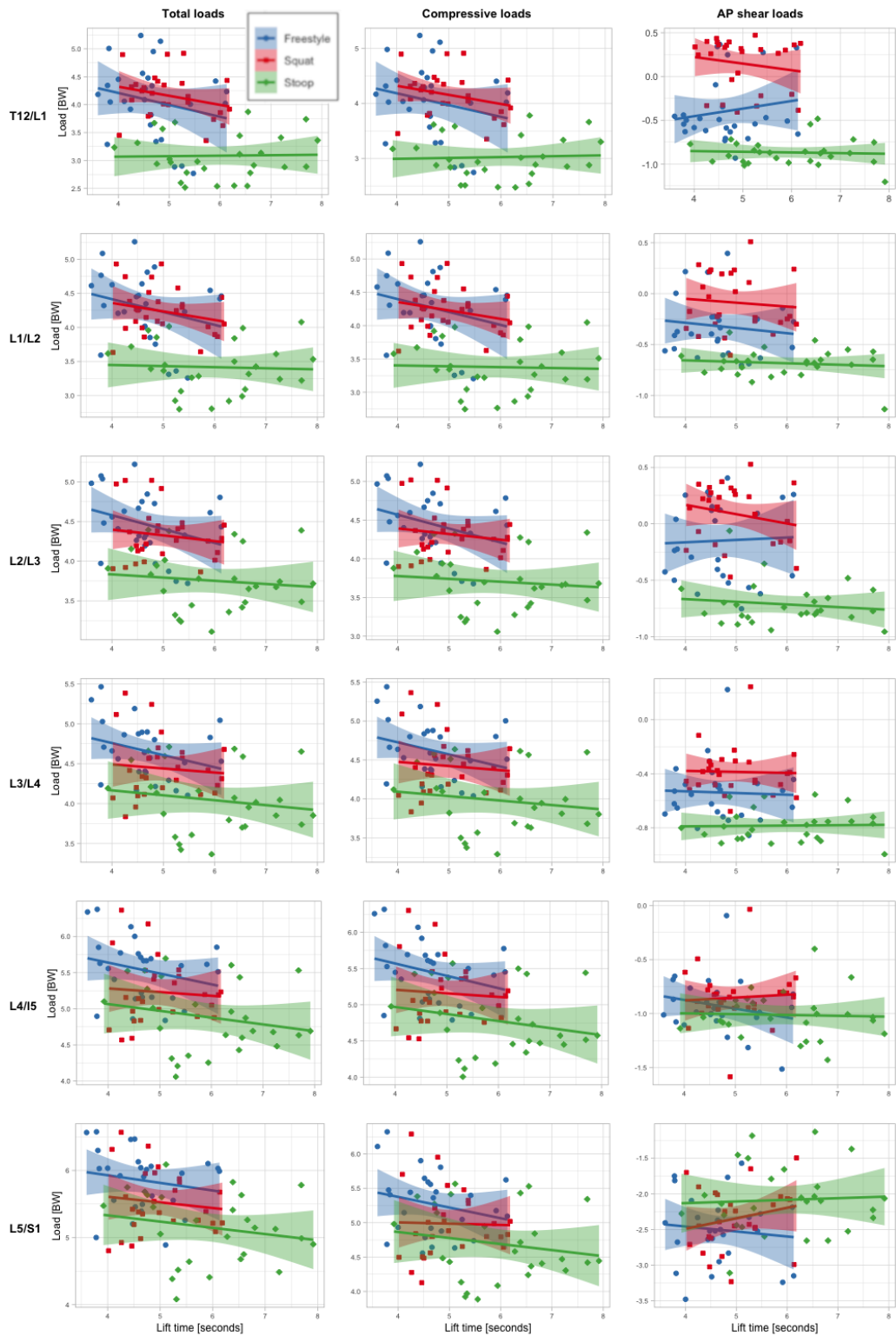

**Supplementary Figure 1**. Scatterplots display specific peak loads in relation to lift time. Lifting style loads are grouped by color. Colored lines indicate regression lines for respective styles, standard error is indicated by the shaded band around the regression line.